\documentclass[article]{aa} 
\usepackage{graphicx}
\usepackage{longtable}
\usepackage{supertabular}
\usepackage{booktabs,multirow,array}
\usepackage{caption}
\usepackage{subcaption}
\usepackage[switch, modulo]{lineno}
\usepackage{txfonts}
\usepackage{xcolor}
\newcommand{\fliper}{{FliPer} }
\newcommand{\fliperp}{{FliPer}}
\newcommand{\numax}{{$\nu_{max}$} }
\newcommand{\numaxp}{{$\nu_{max}$}}

\newcommand{\az}{{A2Z} }

\newcommand{\kepler}{{\textit{Kepler}} }

\newcommand{\re}{\textcolor{black}}
\begin{document} 
   \title{FliPer: A global measure of power density to estimate surface gravities of \re{main-sequence} Solar-like stars \re{and red giants}}
   \titlerunning{FliPer}

   \author{L. Bugnet
          \inst{1,}\inst{2}
          \and R. A. García\inst{1,}\inst{2} 
          \and G. R. Davies\inst{3,}\inst{4} 
          \and S. Mathur\inst{5,}\inst{6,}\inst{7}
          \and E. Corsaro\inst{8}
          \and O. J. Hall\inst{3,}\inst{4} 
          \and B. M. Rendle\inst{3,}\inst{4}}

\institute{IRFU, CEA, Universit\'e Paris-Saclay, F-91191 Gif-sur-Yvette, France\\
\email{lisa.bugnet@cea.fr}
\and
AIM, CEA, CNRS, Université Paris-Saclay, Université Paris Diderot, Sorbonne Paris Cité, F-91191 Gif-sur-Yvette, France
\and 
School of Physics and Astronomy, University of Birmingham, Edgbaston, Birmingham, B15 2TT, UK
\and 
Stellar Astrophysics Centre, Department of Physics and Astronomy, Aarhus University, Ny Munkegade 120, DK-8000 Aarhus C, Denmark
\and
Instituto de Astrof\'{\i}sica de Canarias, E-38200, La Laguna, Tenerife, Spain
\and 
Universidad de La Laguna, Dpto. de Astrof\'{\i}sica, E-38205, La Laguna, Tenerife, Spain
\and 
Space Science Institute, 4750 Walnut Street Suite 205, Boulder, CO 80301, USA
\and
INAF - Osservatorio Astrofisico di Catania, Via S. Sofia 78, I-95123 Catania, Italy}
   \date{Received 27 mars 2018/ Accepted 13 september 2018}


  \abstract
 {Asteroseismology provides global stellar parameters such as masses, radii or surface gravities using the mean global seismic parameters as well as the effective temperature for thousands of low-mass stars ($0.8 M_\odot <M<3 M_\odot$). This methodology \re{has been successfully} applied to stars in which acoustic modes excited by turbulent convection are measured. Other techniques such as the Flicker can also be used to determine stellar surface gravities, but only works for $\log{g}$ above $2.5$ dex. 
In this work, we present a new metric called \fliper (\re{the acronym stands for} Flicker in \re{spectral power density}, \re{in opposition to} the standard Flicker measurement which is computed in the time domain) that is able to extend the range for which reliable surface gravities can be obtained ($0.1<\log{g}<4.6$ dex) without performing any seismic analysis \re{for stars brighter than \textit{Kp} $<$ 14}. \fliper takes into account the average variability of a star measured in the power density spectrum in a given range of frequencies. However, \fliper values calculated on several ranges of frequency are required to better characterize a star. Using a large set of asteroseismic targets it is possible to calibrate the behavior of surface gravity with \fliper through machine learning. This calibration made with a random forest regressor covers a wide range of surface gravities from main-sequence stars to subgiants and red giants, with very small uncertainties from $0.04$ to $0.1$ dex. \fliper values can be inserted in automatic global seismic pipelines to either give an estimation of the stellar surface gravity or to assess the quality of the seismic results by detecting any outliers in the obtained \re{\numax} values. 
\fliper also constrain the surface gravities of main-sequence dwarfs using only long cadence data for which the Nyquist frequency is too low to measure the acoustic-mode properties.}
 

   \maketitle

\section{Introduction}

The precise knowledge of stellar parameters is crucial for a very broad range of fields in astrophysics. Indeed, while it helps us understanding stellar evolution, it also provides important information needed for planetary searches and for studying the chemical and dynamical evolution of our Galaxy. In the last decade, the NASA mission {\it Kepler} \citep{2013Sci...340..587B} collected very high-quality photometric data for almost 200,000 stars \citep{2017ApJS..229...30M} continuously during $\sim$\,4 years. These observations not only revolutionized the search for exoplanets but also opened a window into stellar physics. Asteroseismology proved to be a very powerful tool to better characterize the stars in terms of mass, radii, and age \citep{2010ApJ...723.1583M,2012ApJ...749..152M,2017ApJ...835..173S,2017ApJS..233...23S} but also in terms of their rotation and magnetic activity \citep{2014ApJS..211...24M,2014A&A...572A..34G,2015MNRAS.446.2959D,2017A&A...605A.111C,2017A&A...598A..77K}.  However stellar oscillations have not been detected neither in all red giants \citep[$\sim$\,16,000 reported out of the $\sim$\,24,000 in the latest \emph{Kepler} star-properties catalog]{2017ApJS..229...30M}, nor in all the main-sequence Solar-like stars. Indeed, around 135,000 main-sequence dwarfs have only been observed in long cadence (LC, sampling time of 29.4\,min) by \emph{Kepler} preventing any direct asteroseismic analyses because their acoustic-mode frequencies are well above the Nyquist frequency and can only be seismically studied with short-cadence data \citep[SC, sampling time of 58.85\,s]{2011Sci...332..213C}. \\

To circumvent this, new techniques are being developed to extract precise surface gravities ($\log{g}$) directly from the photometric data. This is the case of the Flicker, i.e., the measurement of the brightness variations in timescales shorter than 8 hrs \citep{2013Natur.500..427B,2016ApJ...818...43B}, the variance of the flux \citep{2012A&A...544A..90H}, the granulation \citep{2011ApJ...741..119M,2014A&A...570A..41K}, and from the analysis of the time scales of convective-driven brightness variations \citep{2016SciA....250654K}. However all these techniques have limitations. Flicker is restricted by construction to stars with $4500<T_{\rm{eff}}<7150$ K and $2.5 < \log{g} < 4.6$ dex, preventing the study of high-luminosity red giant branch (RGB) and asymptotic red giant branch (AGB) stars. To obtain the granulation properties it is necessary to fit a complicated model including different scales of convection with many free parameters \citep[for more details see the discussions in][]{2011ApJ...741..119M,2014A&A...570A..41K,2017A&A...605A...3C}. The final method requires that the oscillation signal is temporally resolved preventing to extend the analysis to main-sequence dwarfs only observed in long-cadence data. It has also been shown that instead of using classic seismic methods it is possible to apply machine learning algorithms directly on the data. For instance, \cite{2018MNRAS.tmp..471H} apply a convolutional neural network on spectra to classify stars. This method gives good results for about $99$~\% of their sample of red giants, including some stars that were not already characterized with seismic pipelines. A Random Forest regression model \re{\citep{Breiman2001}} applied directly on the photometric light curves of variable stellar sources can also estimate their surface gravity with a 0.42 dex uncertainty \citep{2014AAS...22312501M}. \\

We present here a new metric called FliPer (Flicker in Power) --in opposition to the standard Flicker measurement which is computed in the time domain-- that aims at linking the variability of a star to its surface gravity in a wider range than the Flicker, starting at a $\log{g} \sim 0.1$ and similar effective temperatures ($4500<T_{\rm{eff}}<7150 \textnormal{K}$) covering Solar-like pulsating stars. We are limited in the $0.1 < \log{g} < 4.6$ dex range of surface gravity because of the lack of information we have on extreme surface gravity Solar-like stars. There is no intrinsic limits of applicability to the \fliper calculation. We decide to combine powerful methods: we include \fliper values from different lower frequency boundaries into a supervised machine learning Random Forest algorithm in order to get even more accurate results on the surface gravity estimation. This way, we obtain information about the impact of the lower frequency boundaries and the effective temperature on the estimation of surface gravity.

\section{Observations, data selection and preparation}
\label{data}
    In this work, long (29.4 minutes) cadence data \citep{2010ApJ...713L.160G} obtained by the NASA's \emph{Kepler} main mission have been used. The light curves have been corrected and the different quarters concatenated following \cite{2011MNRAS.414L...6G}. Two high-pass filters have been used with cut-off frequencies corresponding to 20 and 80 days. To minimize the effects of the gaps in the observations \citep{2014A&A...568A..10G} the missing observations have been interpolated using impainting techniques \citep{2015A&A...574A..18P}. \re{The power spectrum density is then computed for each star (calibrated as single-sided spectrum)}. Data are corrected from apodization following \cite{2011ApJ...732...54C}.\\
    
    We selected $\sim 15,000$ red-giant stars (RG) among the ones in \cite{2017ApJS..229...30M} showing stellar pulsations and characterized using the \az asteroseismic pipeline \citep{2010A&A...511A..46M}. These stars have $0.1 < \log{g} < 3.4$ dex and $3285 \textnormal{K}<T_{\rm{eff}}< 7411 \textnormal{K}$. \re{I}n addition, $254$ main-sequence (MS) stars with $4951 \textnormal{K} <T_{\rm{eff}}< 6881 \textnormal{K}$ are used to extend the study toward\re{s} higher surface gravity range, reaching $4.5$ dex. \re{These stars have \emph{Kepler} magnitudes brighter than 14 (\textit{Kp}$<14$)}.\\
    
    \re{It is important to notice that the values of \numax computed by A2Z do not show any systematic biases at a level of $\sim$ 1 $\%$ when compared to other seismic pipelines as shown by \citet{2018arXiv180409983P}.}

\section{The new metric: FliPer}

The complete power spectrum contains contributions from the stellar variability at all time scales such as oscillation modes, surface granulation, and rotation. We define \fliper as:
    
    \begin{equation}
        \text{F}_{\textnormal{p}} = \overline{\text{PSD}} - \text{P}_\textnormal{n} \; ,
        \label{powvar}
    \end{equation}
where $\overline{\textsc{psd}}$ represents the averaged value of the power spectrum density from a giving frequency (see Section \ref{range}) to the Nyquist frequency and $\textnormal{P}_\textnormal{n}$ is the photon noise. This noise could be calculated by taking the average value of the \textsc{psd} over a range of frequencies close to the Nyquist frequency, but this method leads to biased estimation of \fliper for stars that oscillate with a frequency close to the Nyquist frequency as explained in detail by \cite{2017sf2a.conf...85B}. \re{Then, the photon noise has been computed following the empirical expression obtained by \cite{2010ApJ...713L.120J}.}\\

The value of \fliper is dominated by a combination of the granulation and the oscillation modes that both depend on the evolutionary stage of the star. The more evolved the star, the larger their oscillation and granulation amplitudes \citep[e.g.][]{2012A&A...537A..30M,GarStello2015}, while the frequency of maximum power $\nu_{\rm{max}}$ decreases \citep[e.g.][]{2014aste.book...60B}.\\

It is important to notice that the signature of strong rotation (and its harmonics) would bias \fliperp. This doesn't have a large impact for the case of red giants because a very small fraction of them shows signatures of the rotation in the PDS as shown by \cite{2017A&A...605A.111C} but needs to be studied in details for main-sequence Solar-like stars (see Section \ref{ROTA}). \\

\subsection{Computing \fliper from data}
\label{range}
The observational frequency range used to compute $\overline{\textsc{psd}}$ is limited at high frequency by the Nyquist frequency. For most stars (those observed in LC) we cannot get information above $\sim 283$ $\mu Hz$. Therefore, we select a first set of calibrator stars including red giant pulsating at a frequency lower than $300$ $\mu Hz$ and for which asteroseismic parameters are available. A second set of known seismic main-sequence dwarfs is used to study \fliper with long cadence data only.\\

     \begin{figure*}[!ht]
            \includegraphics[width=\hsize]{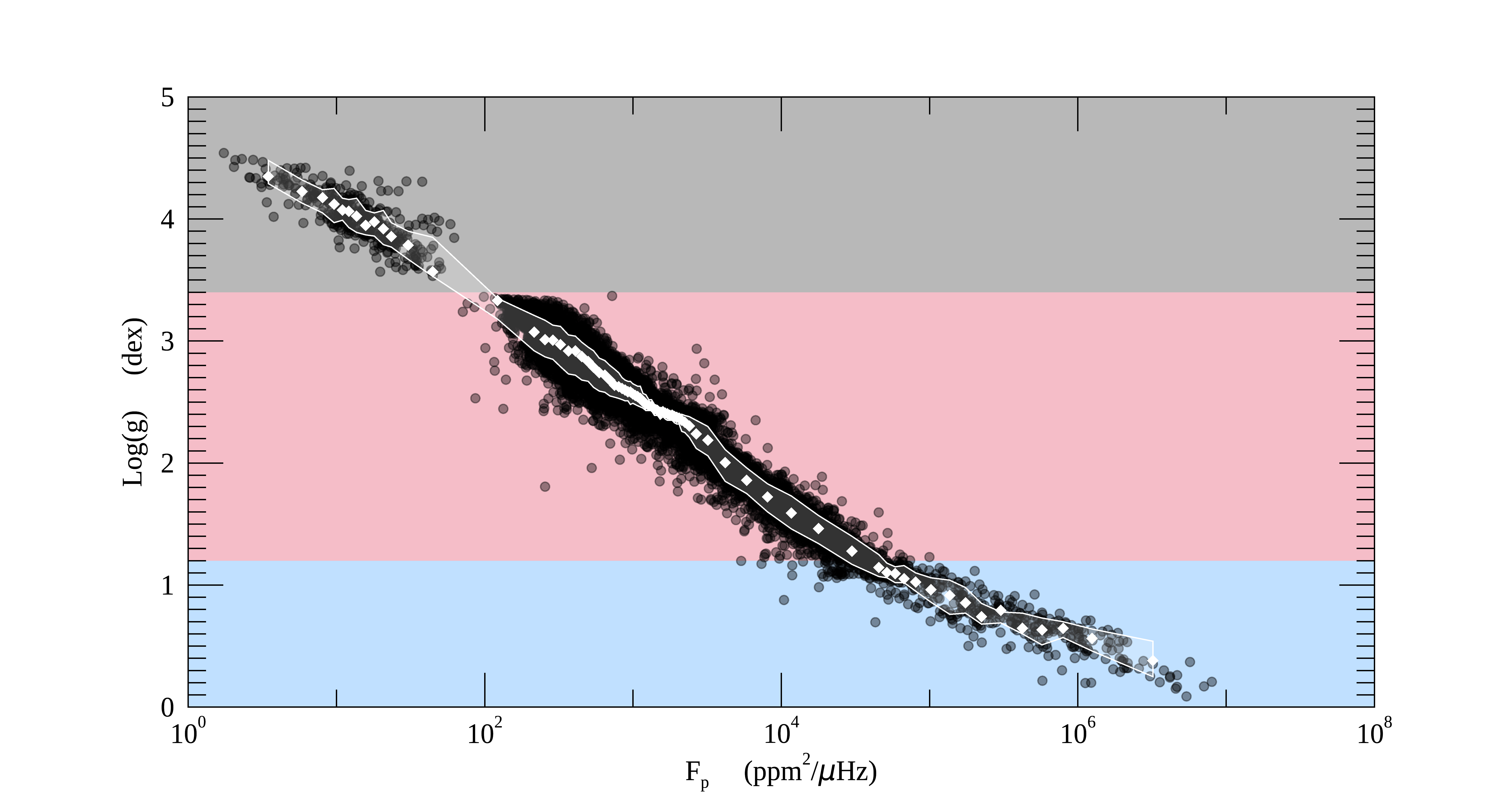}
            \caption{Seismic $\log{g}$ \textsc{vs} \fliper for $\sim 15,000$ stars observed with \textit{Kepler} long-cadence observational mode (black dots). The three shaded regions, grey, red and blue correspond to MS and sub-giant stars, RG, and high-luminosity RG stars or AGB respectively. White diamonds represent the weighted mean value of $\log{g}$ \re{(see Eq.~\ref{mean})} computed with 30 MS stars and 300 RGs each. The white area delimits the location where $68\%$ of the stars in our sample are around the mean value. The white boundary represents the equivalent of a 1$\sigma$ uncertainty (standard deviation) on the surface gravity obtained from \fliperp.}
            \label{HL}
  \end{figure*}
  
The low-frequency limit of $\overline{\textsc{psd}}$ is given by the cut-off frequency used in the calibration of the data. For most of the stars, a 20 days high-pass filter light curves is used. The associated cut-off frequency of the signal is $0.58$ $\mu Hz$. We thus establish a low-frequency limit for the analysis at $0.7$ $\mu Hz$. As main-sequence stars can rotate with a period shorter than 20 days, \fliper is computed with a low-frequency limit at $7$ $\mu Hz$ (i.e. $\sim$1.6 days) avoiding most of the pollution induced by rotation signals (see Section \ref{ROTA}). For stars showing rotation harmonics at higher frequencies, the low-frequency boundary should be taken even higher (e.g. $20$ $\mu$Hz) to avoid any additional impact on \fliper from the peaks associated with rotation. Finally, a small amount of red giants in our sample are either high-luminosity Red Giant Branch stars or AGB stars ($\log{g} < 1.2$ dex) pulsating at frequencies smaller than the 20 days cut-off frequency of the calibrated data. For these stars, an 80 days filter is used in the calibration process. It allows us to properly measure the stellar signal down to $0.2$ $\mu Hz$ (which is the limit frequency utilized in this analysis) and to include oscillation-mode power into the \fliper value.\\

\subsection{A first surface gravity estimator}

For stars with Solar-like oscillations, seismic surface gravities are directly obtained from the frequency of maximum oscillation power \numax computed with the A2Z pipeline \citep{2010A&A...511A..46M} and effective temperatures from the \kepler DR25 catalog \citep{2017ApJS..229...30M}. Knowing seismic surface gravities with their uncertainties allows us to study the behavior of \fliper with the evolutionary state of the stars using only LC light curves even for main-sequence stars. It is important to notice that for main-sequence stars, the seismic $\log{g}$ has been seismically inferred using SC data although \fliper has been computed using LC data.\\

In Fig.~\ref{HL}, the seismic $\log{g}$ is represented as a function of \fliper  (black dots). Three different areas have been identified depending on the evolutionary state of the star: MS and sub-giant stars (grey shaded region), RG (red), and high-luminosity RG stars from the branch and from the asymptotic branch (blue). For each of these category of stars \fliper has been computed with a different low-frequency limit of $7$ $\mu$Hz (avoiding in most cases the region of possible pollution by rotation signatures present on data filtered with a high-pass filter at 20 $\mu$Hz), $0.7$ $\mu$Hz (20 days filter), and $0.2$ $\mu$Hz (80 days filter) respectively. The color scheme is universal in the captions.\\

In order to characterize the relationship between \fliper and $\log{g}$ represented in Fig.~\ref{HL}, we calculate an averaged value of $\log{g}$ for each bin of $n$ stars ($n=300$ for \textsc{RG} and $n=30$ for MS and sub-giant stars) \re{as follows:} 

\begin{equation}
\re{
\overline{\re{\textnormal{log{g}}}}=\frac{\sum_{i=1}^{n} \frac{1}{\delta {\textnormal{log{g}}_{i}}}\times{\textnormal{log{g}}_{i}}}{\sum_{i=1}^{n} \frac{1}{\delta {\textnormal{log{g}}_{i}}}}}		\; ,
\label{mean}
\end{equation}
\re{where $\textnormal{log{g}}_{i}$ represents the value of surface gravity for each star and $\delta \textnormal{log{g}}_{i}$ the corresponding uncertainty.}

These values are represented by the white diamonds, and are located at the averaged value of \fliper over each bin. To define the $1 \sigma$ uncertainties, we compute the area containing 68\% of the stars of the sample (marked by a white contour region in Fig.~\ref{HL}). Mean values and their corresponding $\pm$ 1$\sigma$  uncertainties are reported in Table.~\ref{tab1}. By using these mean values it is possible to estimate the stellar surface gravity directly from the \fliper estimator. The uncertainties obtained on $\log{g}$ extend from $0.05$ to $0.2$ dex, depending on the evolutionary state of the star. \re{Because of the calculation of the mean values, the boundaries in $\log{g}$ are reduced to the range 4.35 to 0.38 dex, as shown in Table~\ref{tab1}.}

\begin{figure*}[!ht]
            \includegraphics[width=1.01\hsize]{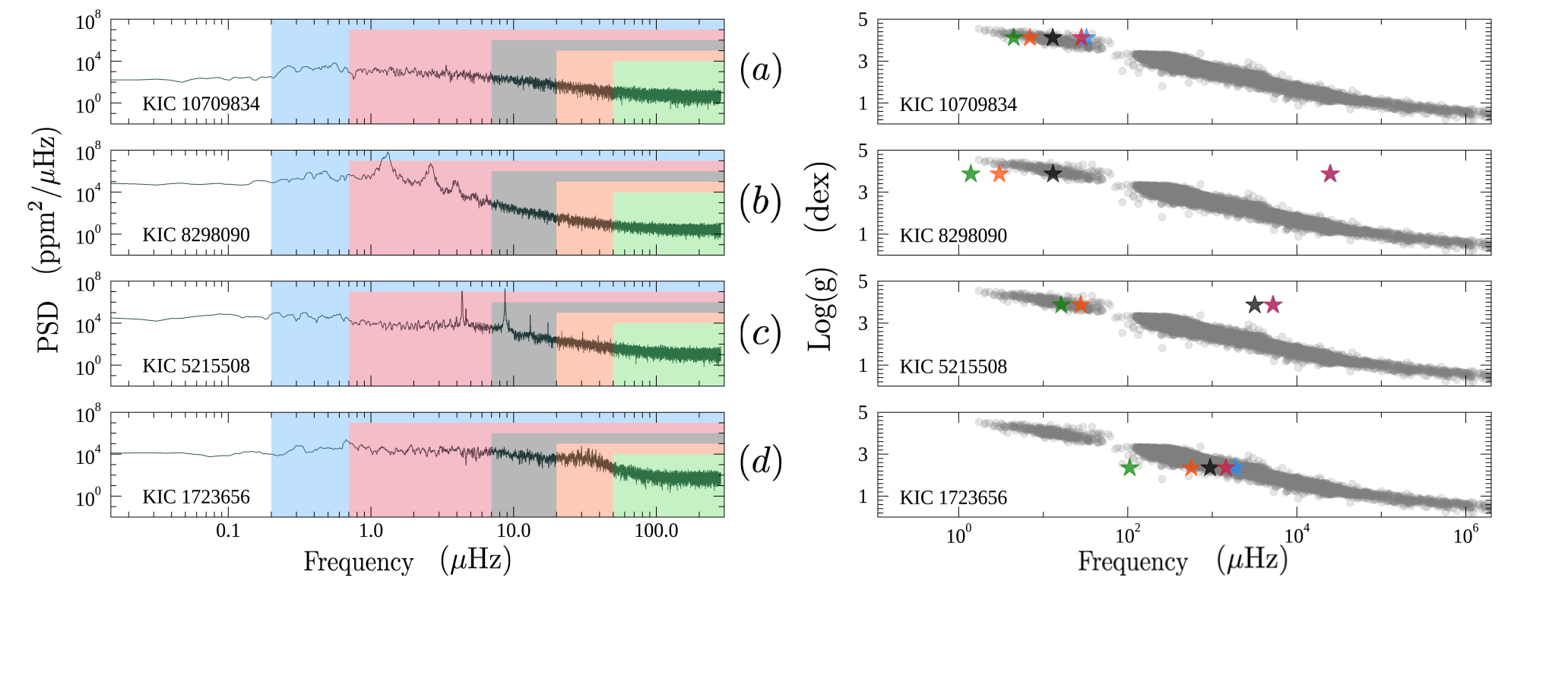}
            \caption{Impact of the lower frequency limit in the \fliper calculation on the estimation of surface gravity for different type of star. \textbf{Left panels:} Power density spectra of four \textit{Kepler} targets. Colored areas (resp. blue, red, black, orange and green) represent the different ranges of frequency used for \fliper calculation (from resp. $0.2$, $0.7$, $7$, $20$ and $50$ $\mu$Hz to the Nyquist frequency). The color scheme is universal in the captions. \textbf{Right panels:} All studied \textit{Kepler} stars from Fig.~\ref{HL} are represented in grey in the $\log{g}$ V.S \fliper diagram. Colored stars (blue, red, black, orange and green) show the position in the diagram of the four stars from left panels with their color corresponding to the low-frequency boundaries used to compute the \fliper value. Panel (a) represents a MS star without any visible rotation component, panel (b) a MS star showing rotation, panel (c) a high frequency rotating MS star, and panel (d) a RG star. }
            \label{ROT}
\end{figure*}

\subsection{Disentangling Main-Sequence stars from Red Giants}

\label{ROTA}

As it is defined, \fliper is mostly dominated by a combination of the power coming from granulation and oscillation modes (when the latter are below the Nyquist frequency). The limitation in the use of the calibrated values from Table~\ref{tabl} to directly estimate surface gravity of stars appears when the spectrum shows a specific behavior that strongly modifies the mean value of the power density. For instance, in stars showing large excess of power (e.g. due to spikes at thrusters frequency in K2 data, or to pollution from a background binary), the value of \fliper is biased towards high power density \citep{2017sf2a.conf...85B}. On the contrary, in stars with a low signal-to-noise ratio the value of \fliper is biased towards lower values, because most of the spectrum is dominated by the instrumental noise. As a consequence, \fliper is higher than expected for fast rotating MS stars due to the rotation peaks and their harmonics, which can be particularly high for young main-sequence dwarfs. For these stars, the  $\log{g}$ inferred from Table~\ref{tabl} could be such that it corresponds to a RG and not to a MS star, even if we calculate \fliper with the $7 \mu$Hz frequency limit. \\

To avoid this problem and to disentangle any MS stars from RGs, we need an additional parameter that takes into account the power due to rotation. The most simple solution is to combine different \fliper values, including some at higher frequencies than the $7 \mu$Hz limit. For each star in our sample, we then calculate \fliper with several low-frequency limits (e.g. F$_{{\textnormal{p}_{0.2}}}$ from 0.2, F$_{{\textnormal{p}_{0.7}}}$ from 0.7, F$_{{\textnormal{p}_{7}}}$ from 7, F$_{{\textnormal{p}_{20}}}$ from 20 and F$_{{\textnormal{p}_{50}}}$ from 50 $\mu Hz$). For MS stars with small rotation signatures the value of \fliper is almost the same for all the low-frequency boundaries (see panel (a) on Fig.~\ref{ROT}). However, when rotation peaks are present, there is a large difference between the \fliper parameters, depending on the frequency of the rotational peaks (See Fig.~\ref{ROT}). This is the case for both stars KIC 8298090 and KIC 5357446 represented on panels (b) and (c). On panel (b) all the rotational components are below the $0.7 \mu$Hz boundary, meaning that parameters F$_{{\textnormal{p}_{20}}}$ and F$_{{\textnormal{p}_{50}}}$ were not necessary to classify this star as a main-sequence star. However on panel (c) the rotation peaks reach higher frequencies: in order to estimate the surface gravity of this star the two new high-frequency parameters are needed. Panel (d) shows a RG star for comparison. In the regime of RG stars, all the \fliper values are very similar - except the lowest ones coming from the calculation with high-pass filter that doesn't include the range of frequency of oscillation modes. By comparing the values of \fliper computed with different low-frequency limits, it is then possible to disentangle MS stars with a high rotation signature from RG stars. \re{This can be done in a star-by-star analysis as it is done in Figure~\ref{ROT}. However, it is possible to automatize this procedure as it is explained in the following section.}

\section{Seismic independent surface gravity prediction from {0.1} to 4.5 dex}
\label{train}
The direct estimation of surface gravity from Table~\ref{tabl} gives good results only when the evolutionary state of the star is already known, and when the spectrum does not show a specific behavior that strongly modifies the mean value of the power density \re{(e.g. when the PSD is polluted by spikes of a background binary or a classical pulsator)}. The reason is that we only use one value of \fliper computed from one lower frequency limit. Estimating surface gravities of unclassified or complex stars requires a different use of the \fliper method.\\

\begin{figure*}[!htb]
            \includegraphics[width=1.0\hsize]{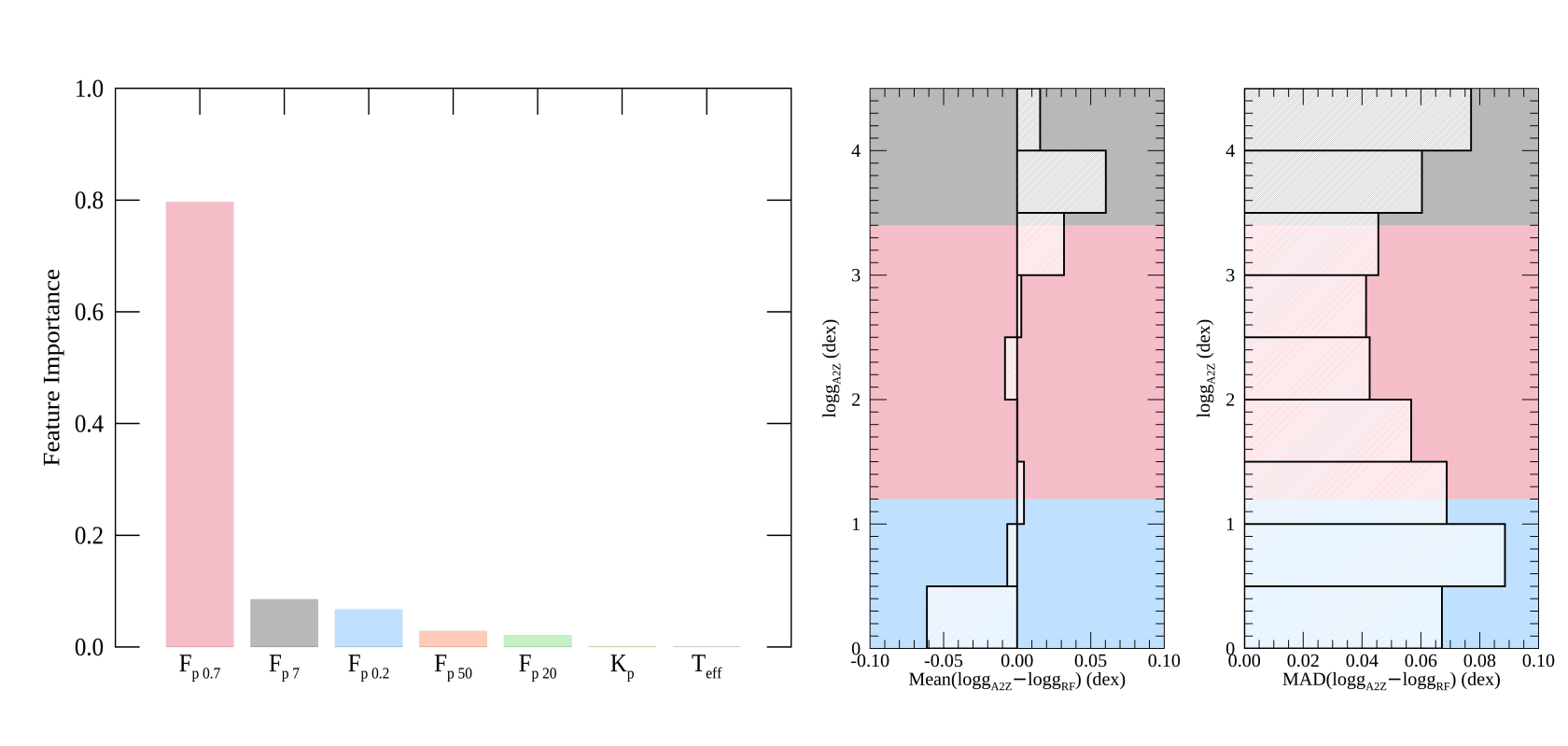}
\caption{\textbf{Left panel:} Importances of the different parameters F$_{{\textnormal{p}_{0.2}}}$, F$_{{\textnormal{p}_{0.7}}}$, F$_{{\textnormal{p}_{7}}}$, F$_{{\textnormal{p}_{20}}}$, F$_{{\textnormal{p}_{50}}}$,T$_{\textnormal{eff}}$ and \textit{Kp} on the training process. The color scheme is universal in the captions. \re{\textbf{Middle panel:} Histogram of the residuals of the estimated $\log{g}_\textnormal{RF}$ values from the references $\log{g}_\textnormal{A2Z}$. \textbf{Right panel:} Histogram of the mean absolute deviation from the expected value MAD($\log{g}_\textnormal{A2Z} - \log{g}_\textnormal{RF}$).\label{importance}}}
\end{figure*}

\re{\subsection{Using machine learning}}
%
As explained above, combining different \fliper values is powerful to detect MS stars showing high rotation signal among RGs. It also means that by using different high-pass filters in the calculation of \fliper we are sensitive to different physical signatures in the PSD. Combining them in the study thus improves the characterization of the star, and we intend to use this wisely to predict surface gravities. To do so, we train a Random Forest regressor algorithm \re{\citep{Breiman2001}} on a random subsample representing \re{8}0\% of our set of stars. The Random forest method is based on the aggregation of a large number of decision trees \re{(see Appendix~\ref{random} for a description of the method) that has already been proved to be useful in asteroseimology \citep[e.g.][]{2014AAS...22312501M}}. The trees are constructed from a training data set and internally validated to give prediction given the predictor for future observations. The random forest method not only allows the use of a large number of parameters but also the estimation of their individual impact on the regression. The parameters used \re{to estimate surface gravities} are F$_{{\textnormal{p}_{0.2}}}$, F$_{{\textnormal{p}_{0.7}}}$, F$_{{\textnormal{p}_{7}}}$, F$_{{\textnormal{p}_{20}}}$, F$_{{\textnormal{p}_{50}}}$, T$_{\textnormal{eff}}$ \re{and \textit{Kp}.} They represent the values of \fliper calculated from a low frequency limit of 0.2, 0.7, 7, 20, 50 $\mu$Hz, the effective temperature \re{and the \textit{Kepler} magnitude of the star.}

\re{\subsection{Building training and test sets of parameters for the random forest algorithm\label{incert}}}

\re{We intend to take into account the uncertainties on the parameters during the testing of the algorithm to estimate the intrinsic bias and/or uncertainties of our methodology. The uncertainties on effective temperature $\delta \textnormal{T}_\textnormal{eff}$ are taken directly from the \cite{2017ApJS..229...30M} catalog. The error on the surface gravity $\delta \textnormal{log{g}}$ comes from the uncertainty on \numax from the seismic analysis of the stars. We can estimate the uncertainty due to the photon noise in the spectra (following a chi-squared distribution with two degrees of freedom) impacting the determination of \fliper by considering negligible the uncertainty made on the photon noise. Hence, the uncertainty on \fliper can be explicitly written as: 
\begin{equation}
\delta \textnormal{F}_\textnormal{p}=\sqrt{\delta \overline{\textnormal{PSD}}^2}=\frac{\delta \textnormal{P}_\textnormal{tot}}{\textnormal{N}_\textnormal{bin}}=\frac{\sqrt{\sum_{i}{\delta \textnormal{P}_i^2}}}{\textnormal{N}_\textnormal{bin}}  \; ,
\end{equation}
where $\delta \textnormal{P}_i$ stands for the error made on the power contained in each bin and $\textnormal{N}_\textnormal{bin}$ is the total number of bins in the power density spectrum. The individual $\delta \textnormal{P}_i$ can not be extracted directly because the $\chi^2_2$ noise distribution does not have gaussian errors. We thus use the central limit theorem and we rebin the spectrum by a factor of $n=50$. The total amount of power in the spectrum is then expressed as:
\begin{equation}
\textnormal{P}_{tot}=\sum_{j}{\textnormal{P}_\textnormal{n,j}} \; ,
\end{equation}
where $\textnormal{P}_\textnormal{n,j}$ follows a quasi normal distribution with $2\textnormal{n}$ degrees of freedom. This of course assumes that the signal does not change dramatically over this range of $50$ bins, which is consistent with the shape of the spectra in solar-like stars. The uncertainty on the mean from each $n$ bins is then expressed as:
\begin{equation}
\delta \textnormal{P}_\textnormal{n,j}=2\times \frac{\textnormal{P}_\textnormal{n,j}}{2n} \times \sqrt{n}
\end{equation}
leading to a global uncertainty on \fliper values of 
\begin{equation}
\delta \textnormal{F}_\textnormal{p}=\frac{\sqrt{\sum_{j} \left( 2\times \frac{\textnormal{P}_\textnormal{50,j}}{2 n} \times \sqrt{n} \right)^2 }}{\textnormal{N}_\textnormal{bin}} \; .
\end{equation}
 \\
Then, we include the effect of these errors on the different parameters on the testing of the algorithm. To do so, we perform a Monte Carlo simulation by generating for each star in our test sample (representing $20\%$ of the total amount of stars in our study) $100$ artificial sets of parameters from their corresponding normal distributions. With $\mathcal{G}_{0\le i \le 100}$ being $100$ random values following the standard distribution, we calculate for each $\textnormal{X}$ parameter (F$_{{\textnormal{p}_{0.2}}}$, F$_{{\textnormal{p}_{0.7}}}$, F$_{{\textnormal{p}_{7}}}$, F$_{{\textnormal{p}_{20}}}$, F$_{{\textnormal{p}_{50}}}$,T$_{\textnormal{eff}}$ and $\log{g}$) $100$ new values $\textnormal{X}_{0\le i \le 100}$ following Eq.~\ref{variable} below. However, the $Kepler$ magnitude of the star remains constant as it has no uncertainties, and it completes each of the 100 new sets of parameters.
\begin{equation}
\textnormal{X}_{0\le i \le 100}=\textnormal{X}+\delta \textnormal{X} \times \mathcal{G}_{0\le i \le 100} 
\label{variable}
\end{equation}
}
\re{Each new group of 100 sets of parameters is considered in the following study as a representation of a hundred stars to test the algorithm.}

\re{\subsection{Impacts of parameters on the training}}
\re{We use $80\%$ of our stars to train the algorithm to estimate the surface gravity. The remaining $20\%$ of stars is used to test the performance of the algorithm, by taking into account uncertainties on the different parameters as explained in Section~\ref{incert}. The impacts of the different parameters on the training process are represented on Fig.~\ref{importance} A.}\\

A predictable result is that the F$_{{\textnormal{p}_{0.7}}}$ parameter largely dominates the training. It comes from the fact that this is the most suitable parameter to study RG, representing more than $90 \%$ of the total amount of stars. Other relevant values of the filtering appear to be 7 and 0.2 $\mu$ Hz. Indeed, F$_{{\textnormal{p}_{7}}}$ plays an important role in the training because of its ability to distinguish MS stars from RGs, and the F$_{{\textnormal{p}_{0.2}}}$ parameter helps in the prediction of surface gravity for high luminosity stars. The other parameters F$_{{\textnormal{p}_{20}}}$ and F$_{{\textnormal{p}_{50}}}$ have lower impacts on the training, but still help the learning for high rotating MS stars. \re{Impacts of the effective temperature and \textit{Kp} do not exceed a few percent}. We confirm from Fig.~\ref{importance} that combining different lower-frequency boundaries in the \fliper calculation makes a great difference for the estimation of robust surface gravities. \\

\re{\subsection{Results}}
To evaluate the performance of the algorithm, the estimate of surface gravity from the test sample is compared to the corresponding \az estimation of surface gravity. The mean \re{absolute} 
deviation (MAD) of the Random Forest surface gravities from reference values is reported in Table~\ref{param1}. This estimator of the deviation is chosen to be robust against outliers to avoid any issue coming from an eventual remaining error in the A2Z estimation of surface gravity.\\

The estimation of surface gravity resulting from the machine learning on the test sample has an averaged deviation of $\sim 0.0\re{46}$ dex from our reference values (see Table~\ref{param1}). We also get surface gravity deviation and errors from the reference values for different ranges of surface gravity on Table~\ref{param1}. We conclude that for all our stars, the new method gives a very good precision on surface gravity: the Flicker method in the range of $\log{g}$ of $2.5-4.6$ dex have typical errors between $0.1$ and $0.2$ dex. \re{Here, errors lie from $\sim 0.04$ dex for red giants to $\sim 0.09$ dex for high-luminosity stars (see Fig.~\ref{importance} right panel). Our estimates are in average centered around the $\log{g}$ reference values (see Table~\ref{param1}). There is a small bias for HL and MS stars (see Fig.~\ref{importance} middle panel) because the estimation of extreme surface gravities is the hardest for the algorithm, which sometimes gets a little biased by the presence of lots of red giants in the sample. This bias that depends on the evolutionary state of the star should be taken into account, but it remains smaller than the uncertainties on the original surface gravity values.}\\

\re{Our algorithms are available on GitHub (\url{https://github.com/lbugnet/FLIPER}) where the functions to calculate \fliper and the Random Forest algorithm are provided. Along with them, we also provide the already trained algorithms for the estimation of surface gravities. They can be directly applied to any solar-type star to estimate its surface gravity from $0.1$ to $4.6$ dex.}\\

\section{Discussion \& Conclusion}

In this work we present a new method to estimate surface gravity of Solar-like stars that extracts information from global power in their spectra. The sample of $\sim 15,000$ stars is constituted of main-sequence and sub-giant stars, stars on the red giant branch (RGB) and clump stars, and also high luminosity stars on the asymptotic giant branch (AGB). This way, we study stars with $0.1 < \log{g} < 4.5$ dex in which mode oscillations are expected to arise from surface convection. Power spectra should then present patterns of granulation power, rotation components, and oscillation-mode power.\\

\addtocounter{table}{0}
\begin{table}
\re{\caption{\re{Summary statistical results on the test set from Fig.~\ref{importance}. MAD is the mean absolute
deviation. }
\label{param1}}             
\centering
\begin{tabular}{@{} l*{2}{>{$}c<{$}} @{}}   
   \toprule
$\log{g}$ range (dex) & \multicolumn{1}{c@{}}{$\overline{\log{g}_\textnormal{A2Z} - \log{g}_\textnormal{RF}}$ (dex)} & \multicolumn{1}{c@{}}{MAD (dex)}\\   
\hline
 & & \\
$\textnormal{ALL}$ & -4.5\times 10^{-4}  & 0.046 \\
 & & \\
$[0 - 0.5]$&	-0.061&     0.067\\
$[0.5 - 1]$&   -0.007		&0.089\\
$[1 - 1.5]$&    0.005&     0.069\\
$[1.5 - 2]$ &   0.000&     0.057\\
$[2 - 2.5]$ &   -0.008&     0.043\\
$[2.5 - 3]$&    0.003&     0.046\\
$[3 - 3.5]$ &     0.032&     0.041\\
$[3.5 - 4]$ &    0.060&     0.060\\
$[4 - 4.5]$ &     0.016 &     0.077\\
\end{tabular}}

\end{table}

\fliper values are calculated by taking the average power density normalized by the photon noise of the star from different lower frequency limits to the Nyquist frequency. Our first method consists of calibrating surface gravity of stars from their \fliper value with a $1 \sigma$ uncertainty (see Table~\ref{tabl}). We explained how these values can be used directly to give a first estimate of surface gravity, however it works well only on stars that are already characterized. Indeed, the evolutionary state has to be known or the star must have a weak rotational signature in order to distinguish main-sequence stars from red giants. To give estimations of surface gravities for any star, we introduce a second method. A Random Forest regressor algorithm is trained to estimate surface gravity on a sample of our stars. We use \fliper values computed with different frequency ranges, spectroscopic effective temperatures and seismic surface gravities. This way, stars are better characterized during the process, and no additional information is needed to provide accurate estimation of surface gravity, even for highly rotating MS stars. By testing the algorithm on the rest of our sample, we obtain estimates of surface gravity with a mean \re{absolute} 
deviation of \re{0.046} dex from seismic $\log{g}$. The training relies on seismic observations of Solar-like stars representing $80$ \% of our sample. However, there is no need for additional seismic measurements to obtain precise estimations of surface gravities on the test set of stars. 
The uncertainty on our results largely improves upon previous non-seismic estimations of surface gravity. Indeed, spectroscopic estimations are known to have \re{0.1-0.3 dex error bars \citep{2016A&A...594A..39F, 2016AJ....151..144G}}. \re{Recent methods such as the Flicker \citep{2016ApJ...818...43B} gives estimates with errors higher than 0.1 dex, while the study of the granulation time-scale \citep{2016SciA....250654K} is limited to stars showing a visible oscillation pattern but with better uncertainties, around 0.018 dex.} In addition, \fliper is extended to a wider range of surface gravities, reaching $\log{g} $ as small as 0.\re{1} dex with a mean \re{absolute} 
deviation of \re{0.046} dex comparable to the other RGs.\\

For main-sequence stars that oscillate at high frequency (above the \emph{Kepler} LC Nyquist frequency), \fliper computed from LC data does not contain mode power, but only granulation-related power \citep{2017A&A...605A...3C} and rotation signals. However, Fig.~\ref{HL} clearly shows that \fliper values for main-sequence stars are still correlated with surface gravity. 
This is a new evidence of the link between granulation and asteroseismic properties \citep{2011ApJ...741..119M, 2014A&A...570A..41K}, allowing us to estimate \numax or rather surface gravities on LC data for which high-frequency modes are not measured. Thus, proper surface gravities can be precisely inferred for any \emph{Kepler} LC solar-like target, from main-sequence to high-luminosity stars, without using direct seismic analysis. \\

\re{Lots of studies concern the estimation of seismic parameters of stars with new techniques directly from the properties of the time series or the power spectrum density \citep[][Keaton Bell et al. submitted]{2017sf2a.conf...85B, 2018ApJ...859...64H,2018MNRAS.tmp.1784P}. We thus adapt our methodology to estimate \numax instead of the surface gravity based on the same sample of stars (see Appendix~\ref{numax}). The results are of course very similar to those on the surface gravity, with uncertainties on \numax about $0.044$ dex and a mean distance to references \numax values ($\sim 1.3 \times 10^{-3}$ dex) negligible. Moreover, FliPer has already been included by Bell et al. (submitted) as a validation procedure to their seismic results and it is also being implemented as one of the parameters to be used in the  classification algorithm that is being developed for the NASA TESS mission \citep{2014SPIE.9143E..20R} using a random forest classifier (Tkachenko et al., in preparation). 
}\\

\begin{acknowledgements}
We would like to thank M. Benbakoura for his help in analyzing the light curves of several binary systems included in our set of stars.
L.B. and R.A.G. acknowledge the support from PLATO and GOLF CNES grants.
S.M. acknowledge support by the National Aeronautics and Space Administration under Grant NNX15AF13G, by the National Science Foundation grant AST-1411685 and the Ramon y Cajal fellowship number RYC-2015-17697.
E.C. is funded by the European Union’s Horizon 2020 research and innovation programme under the Marie Sklodowska-Curie grant agreement no. 664931. O.J.H and B.M.R. acknowledge the support of the UK Science and Technology Facilities Council (STFC). Funding for the Stellar Astrophysics Centre is provided by the Danish National Research Foundation (Grant DNRF106). This research has made use of NASA’s Astrophysics Data System. Some/all of the data presented in this paper were obtained from the Mikulski Archive for Space Telescopes (MAST). STScI is operated by the Association of Universities for Research in Astronomy, Inc., under NASA contract NAS5-26555.
\end{acknowledgements}

\bibliographystyle{aa}

 
\newpage

\newpage

\begin{appendix}
\section{}

\addtocounter{table}{0}
\begin{center}
\tablefirsthead{\hline \hline $\log(\textnormal{F}_\textnormal{P})$ & $\log{g}$ (dex) & -1$\sigma$ (dex) & +1$\sigma$ (dex) \\ } \tablehead{$\log(\textnormal{F}_\textnormal{P})$ & $\log{g}$ (dex) & -1$\sigma$ (dex) & +1$\sigma$ (dex) \\ \hline ... &...&...&...\\  }
\tabletail{... &...&...&...\\}
\tablelasttail{\hline}
\tablecaption{Weighted mean value of $\log{g}$ from diamonds on Fig.~\ref{HL} with their $1 \sigma$ uncertainties for each bin of $30$ (for MS and HL stars) or $300$ (for RGs) stars.\label{tab1}}
\label{tabl}
\begin{supertabular}{c c c c}

\hline
   
        0.54   &          4.35   &          0.06   &          0.13   \\
        0.77   &          4.23   &          0.09   &          0.09   \\
        0.91   &          4.18   &          0.12   &          0.06   \\
        0.98   &          4.12   &          0.15   &          0.13   \\
        1.04   &          4.07   &          0.09   &          0.10   \\
        1.08   &          4.06   &          0.14   &          0.10   \\
        1.14   &          4.02   &          0.13   &          0.14   \\
        1.20   &          3.95   &          0.08   &          0.12   \\
        1.25   &          3.98   &          0.12   &          0.08   \\
        1.32   &          3.92   &          0.13   &          0.15   \\
        1.37   &          3.86   &          0.09   &          0.11   \\
        1.48   &          3.79   &          0.12   &          0.11   \\
        1.65   &          3.57   &          0.03   &          0.28   \\
        2.09   &          3.33   &          0.15   &          0.02   \\
        2.33   &          3.07   &          0.15   &          0.13   \\
        2.41   &          3.01   &          0.14   &          0.16   \\
        2.46   &          3.01   &          0.16   &          0.13   \\
        2.51   &          2.97   &          0.18   &          0.15   \\
        2.56   &          2.92   &          0.19   &          0.14   \\
        2.61   &          2.92   &          0.20   &          0.13   \\
        2.65   &          2.87   &          0.19   &          0.12   \\
        2.70   &          2.83   &          0.16   &          0.12   \\
        2.73   &          2.78   &          0.16   &          0.13   \\
        2.77   &          2.74   &          0.15   &          0.12   \\
        2.81   &          2.72   &          0.14   &          0.12   \\
        2.84   &          2.68   &          0.13   &          0.12   \\
        2.87   &          2.64   &          0.10   &          0.13   \\
        2.90   &          2.63   &          0.09   &          0.11   \\
        2.93   &          2.61   &          0.09   &          0.09   \\
        2.95   &          2.60   &          0.09   &          0.08   \\
        2.97   &          2.58   &          0.08   &          0.09   \\
        2.98   &          2.58   &          0.10   &          0.09   \\
        3.00   &          2.57   &          0.08   &          0.08   \\
        3.01   &          2.55   &          0.08   &          0.09   \\
        3.03   &          2.54   &          0.08   &          0.08   \\
        3.05   &          2.53   &          0.07   &          0.10   \\
        3.07   &          2.50   &          0.05   &          0.07   \\
        3.09   &          2.48   &          0.06   &          0.08   \\
        3.11   &          2.47   &          0.04   &          0.05   \\
        3.12   &          2.46   &          0.04   &          0.05   \\
        3.14   &          2.44   &          0.05   &          0.04   \\
        3.17   &          2.43   &          0.04   &          0.04   \\
        3.18   &          2.42   &          0.05   &          0.04   \\
        3.20   &          2.42   &          0.04   &          0.03   \\
        3.22   &          2.41   &          0.05   &          0.04   \\
        3.23   &          2.40   &          0.05   &          0.04   \\
        3.25   &          2.39   &          0.04   &          0.03   \\
        3.26   &          2.39   &          0.04   &          0.03   \\
        3.28   &          2.38   &          0.05   &          0.04   \\
        3.29   &          2.37   &          0.06   &          0.04   \\
        3.31   &          2.37   &          0.05   &          0.04   \\
        3.33   &          2.35   &          0.09   &          0.05   \\
        3.35   &          2.33   &          0.09   &          0.05   \\
        3.38   &          2.30   &          0.09   &          0.07   \\
        3.43   &          2.24   &          0.12   &          0.11   \\
        3.50   &          2.19   &          0.13   &          0.11   \\
        3.62   &          2.00   &          0.15   &          0.10   \\
        3.77   &          1.86   &          0.11   &          0.10   \\
        3.91   &          1.72   &          0.12   &          0.11   \\
        4.07   &          1.59   &          0.13   &          0.14   \\
        4.25   &          1.46   &          0.12   &          0.11   \\
        4.48   &          1.28   &          0.11   &          0.13   \\
        4.66   &          1.14   &          0.07   &          0.10   \\
        4.71   &          1.10   &          0.04   &          0.08   \\
        4.77   &          1.09   &          0.07   &          0.06   \\
        4.83   &          1.05   &          0.03   &          0.11   \\
        4.91   &          1.02   &          0.08   &          0.08   \\
        5.01   &          0.96   &          0.10   &          0.10   \\
        5.14   &          0.92   &          0.15   &          0.13   \\
        5.24   &          0.85   &          0.08   &          0.13   \\
        5.35   &          0.74   &          0.06   &          0.11   \\
        5.48   &          0.79   &          0.10   &          0.01   \\
        5.63   &          0.65   &          0.04   &          0.12   \\
        5.76   &          0.63   &          0.12   &          0.09   \\
        5.90   &          0.65   &          0.08   &          0.05   \\
        6.10   &          0.56   &          0.10   &          0.08   \\
        6.51   &          0.38   &          0.13   &          0.16   \\

    \hline
\end{supertabular}
\end{center}

\re{\section{Random Forest regressor \label{random}}}

\re{\subsection{Supervised machine learning}}
\re{What is called a Random Forest algorithm is a supervised machine learning (ML) method \citep{Kotsiantis:2007:SML:1566770.1566773}. It learns how to predict an output variable ($\textnormal{Y}_{\textnormal{predicted}}$) from some training data (X) for which the corresponding result ($\textnormal{Y}_{\textnormal{known}}$) is already known. It thus learns a mapping function \textit{f} from the input(s) to the output:
\begin{equation}
\textnormal{Y}_{\textnormal{predicted}}=\textit{f}(X)
\end{equation}
The algorithm iteratively makes predictions ($\textnormal{Y}_{\textnormal{predicted}}$) on the training data (X). They are corrected to achieve a maximum level of performance, by comparing with the $\textnormal{Y}_{\textnormal{known}}$ values.
We use a surpervised ML algorithm for our study because we have input variables X (which are F$_{{\textnormal{p}_{0.2}}}$, F$_{{\textnormal{p}_{0.7}}}$, F$_{{\textnormal{p}_{7}}}$, F$_{{\textnormal{p}_{20}}}$, F$_{{\textnormal{p}_{50}}}$, T$_{\textnormal{eff}}$ and \textit{Kp}) and an output variable $\textnormal{Y}_\textnormal{known}$ (representing our surface gravity $\log{g}$).}

\begin{figure}[h]
            \includegraphics[width=1\hsize]{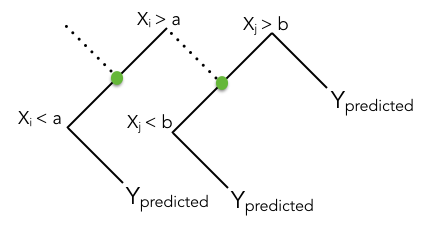} 
\caption{\re{A schematic representation of a regression decision tree. At each node (green points) one variable is split at a value such that the cost function (see Eq~\ref{cost}) is minimized.}}
\label{rf}
\end{figure}

\re{\subsection{Regression trees}}
\re{The regression tree method is part of the Classification and Regression Trees (CART) introduced by \cite{Breiman2001}. A decision tree algorithm constructs a binary tree during the training, with each node representing a split point on a single input variable (X) (a numerical value for regression algorithms, or a class name for classification algorithms). The leaf nodes of the tree contain the output possible predictions ($\textnormal{Y}_\textnormal{predicted}$), as shown in Fig~\ref{rf}.}

\re{The tree is built in such way that the cost function is minimized. Equation~\ref{cost} is the cost function used for the regressor, with $N_{training}$ being the number of stars in our training sample. }

\re{
\begin{equation}
\label{cost}
\textnormal{cost}=\sum_{i=1}^{N_{training}}(\textnormal{Y}_\textnormal{known}-\textnormal{Y}_\textnormal{predicted})^2 \; .
\end{equation}
Once the tree is build on the training sample, it is used to evaluate $\textnormal{Y}_\textnormal{predicted}$ for new $\textnormal{X}_\textnormal{new}$ data.}

\re{\subsection{Ensemble method Random Forest regressor}}
\re{An ensemble method combines the prediction from multiple ML algorithms together. It aims at making even more accurate predictions than any individual model. The Random Forest regressor is an ensemble method that combines regression trees. It consists in:
\begin{itemize}
\item{Creating many sub-samples of the training sample.}
\item{Training a regression tree on each sub-sample, with keeping a low number of variables that can be looked at for each split point. It aims at decreasing the correlation between the different trees. For regression algorithm, the typical number of features that can be searched is $m=\frac{p}{3}$ with $p$ the number of input (X) variables.}
\item{Calculating the average prediction from each model for the new test sample: this averaged value is taken as the estimate for the output variable ($\textnormal{Y}_\textnormal{predicted}$).}
\end{itemize}}

\re{In our work we use the "RandomForestRegressor" function from the "sklearn.ensemble" Python library \citep{scikit-learn} to compute the training on surface gravity.}

\re{\section{An automatic estimation of \numax \label{numax}}}

\re{As a complementary study we also trained our algorithm to estimate the frequency of maximum power \numaxp. The training is made following Section~\ref{train} by using \numax instead of $\log{g}$ as the predicted parameter $\textnormal{Y}_\textnormal{predicted}$. The training input values are computed as in Section~\ref{incert} by combining the \numax values estimated by the \az global seismic pipeline for our sample of $\sim 15,000$ stars along with their uncertainties.}\\

     \addtocounter{table}{2}
\begin{table}
\re{\caption{\re{Summary statistical results on \numax on the test set from Fig.~\ref{importance}. MAD is the mean absolute
deviation. \label{figurea1}}
}             
\centering
\begin{tabular}{@{} l*{2}{>{$}c<{$}} @{}}   
   \toprule
log(\numax) range (dex) & \multicolumn{1}{c@{}}{$\overline{\log{(\nu_{max}}_\textnormal{A2Z}) - \log{(\nu_{max}}_\textnormal{RF})}$ (dex)} & \multicolumn{1}{c@{}}{MAD (dex)}\\   
\hline
 & & \\
$\textnormal{ALL}$ & -0.3\times 10^{-3}  & 0.044 \\
 & & \\
$[-1 : -0.5]$&	-0.123&     0.085\\
$[-0.5 : 0]$&   -0.011		&0.081\\
$[0 : 0.5]$&    -0.0007&     0.075\\
$[0.5 : 1]$ &   -0.004&     0.053\\
$[1 : 1.5]$ &   -0.019&     0.041\\
$[1.5 : 2]$&    0.003&     0.037\\
$[2 : 2.5]$ &     0.025&     0.046\\
$[2.5 : 3]$ &    -0.041&     0.061\\
$[3 : 3.5]$ &     0.011 &     0.082\\
\end{tabular}}

\end{table}

\re{Results are very similar to the estimation of surface gravity, and are given on Table~\ref{figurea1}. The estimation of \numax can be made for any star with solar-like oscillations with $0.1 < \log{g} < 3.4$ dex, $3285<T_{\rm{eff}}< 7411 \textnormal{K}$, \textit{Kp}$ < 14$, and provide a very good prior for any more complex seismic analysis of the star. The complete algorithm for the \numax estimation can be found on the Git repository \url{https://github.com/lbugnet/FLIPER}.}


\end{appendix}

\end{document}